%% file: main.tex
\documentclass[twoside,journey]{IEEEtran}
\pdfoutput=1
\usepackage{ifpdf}
\usepackage{makecell}
\usepackage{hyperref}
\usepackage{array}
\usepackage{graphicx,amssymb,amsmath}
\usepackage{multicol}
\usepackage[noadjust]{cite}
\usepackage{setspace}
\usepackage{graphicx}
\usepackage{float}
\usepackage {url}
\usepackage{stfloats}
\usepackage{amsthm,pifont}
\usepackage{flushend}
\usepackage{cases,subeqnarray}
\usepackage{bm,multirow,bigstrut}
\usepackage{amsmath, amsthm, amssymb}
\usepackage{textcomp}
\usepackage{latexsym,bm}
\usepackage{booktabs}
\usepackage{xcolor}
\usepackage{mathtools}
\usepackage{dsfont}
\usepackage{extarrows}
\usepackage{epsfig}
\usepackage{epsfig}
\usepackage{epstopdf}
\usepackage[noend]{algpseudocode}
\usepackage{algorithmicx,algorithm}
\usepackage{makecell}
\usepackage{colortbl}
\usepackage{booktabs}
\usepackage{graphicx} 
\usepackage{array}    
\usepackage{hyperref} 
\usepackage{booktabs}
\usepackage{enumitem}
\usepackage{diagbox}
\usepackage{tikz}
\usepackage{subcaption}
\usetikzlibrary{trees,shadows.blur}
\usepackage{forest}
\usepackage{caption}
\usepackage{epstopdf}
\theoremstyle{plain}

\theoremstyle{plain}

\usepackage{amsmath}

\newcommand{\ignore}[1]{{{\color{yellow} }}}
\definecolor{blue-green}{rgb}{0.0, 0.87, 0.87}
\IEEEoverridecommandlockouts
\begin{document}

\title{Enhancing Physical Layer Communication Security through Generative AI with Mixture of Experts}

\author{Changyuan Zhao, Hongyang Du, Dusit Niyato, \textit{Fellow, IEEE}, Jiawen Kang, Zehui Xiong, Dong In Kim, \textit{Fellow, IEEE}, Xuemin (Sherman) Shen, \textit{Fellow, IEEE}, and Khaled B. Letaief, \textit{Fellow, IEEE}
\thanks{C. Zhao is with the College of Computing and Data Science, Nanyang Technological University, Singapore, and CNRS@CREATE, 1 Create Way, 08-01 Create Tower, Singapore 138602 (e-mail: zhao0441@e.ntu.edu.sg).
}
\thanks{H. Du, and D. Niyato are with the College of Computing and Data Science, Nanyang Technological University, Singapore (e-mail: hongyang001@e.ntu.edu.sg; dniyato@ntu.edu.sg).}
\thanks{J. Kang is with the School of Automation, Guangdong University of Technology, China. (e-mail: kavinkang@gdut.edu.cn).
}
\thanks{Z. Xiong is with the Pillar of Information Systems Technology and Design, Singapore University of Technology and Design, Singapore (e-mail: zehui xiong@sutd.edu.sg).}
\thanks{D. I. Kim is with the Department of Electrical and Computer Engineering, Sungkyunkwan University, Suwon 16419, South Korea (email:dikim@skku.ac.kr).}
\thanks{X. Shen is with the Department of Electrical and Computer Engineering, University of Waterloo, Canada (e-mail: sshen@uwaterloo.ca).
}
\thanks{Khaled B. Letaief is with the Department of Electrical and Computer Engineering, Hong Kong University of Science and Technology, Hong Kong (e-mail: eekhaled@ust.hk).
}
}

\maketitle
\vspace{-1cm}

\begin{abstract}


AI technologies have become more widely adopted in wireless communications. As an emerging type of AI technologies, the generative artificial intelligence (GAI) gains lots of attention in communication security. Due to its powerful learning ability, GAI models have demonstrated superiority over conventional AI methods. However, GAI still has several limitations, including high computational complexity and limited adaptability. Mixture of Experts (MoE), which uses multiple expert models for prediction through a gate mechanism, proposes possible solutions. Firstly, we review GAI model’s applications in physical layer communication security, discuss limitations, and explore how MoE can help GAI overcome these limitations. Furthermore, we propose an MoE-enabled GAI framework for network optimization problems for communication security. To demonstrate the framework’s effectiveness, we provide a case study in a cooperative friendly jamming scenario. The experimental results show that the MoE-enabled framework effectively assists the GAI algorithm, solves its limitations, and enhances communication security.

\end{abstract}
\begin{IEEEkeywords}

Physical Layer Communication Security, Generative AI, Mixture of Experts, Network Optimization.

\end{IEEEkeywords}
\IEEEpeerreviewmaketitle

\input{Introduction}
\input{Section2}

\input{Section3}

\input{Section4}

\input{Section5}
\input{Conclusion}

\bibliography{Ref}

\end{document}

%% file: Introduction.tex
\section{Introduction}\label{intro}



As communication devices become a part of people's lives, carrying vast amounts of crucial information, the security and privacy of these communications face significant threats. 
Several threats exist in the physical layer of communication systems. The physical layer is a fundamental layer responsible for the raw transmission of bit streams across various transmission media. Attacks on this layer aim to disrupt data transmission or intercept data being transmitted. These vulnerabilities include attacks on communication signals, equipment, and sensors, highlighting the wide range of weaknesses that must be addressed for comprehensive communication network security. 
Due to the foundational level of wireless communication systems,
ensuring the security of the physical layer is the first line of defense in wireless communications.

With the rise of AI technology, there has been a significant shift towards using advanced AI algorithms including machine learning and deep learning to solve security concerns at the physical layer. For example, Convolutional Neural Networks (CNNs) have been applied to facilitate multi-user authentication \cite{liao2019novel}, and Recurrent Neural Networks (RNNs) have 
been utilized to design secure channel coding \cite{xiao2020designing}.
However, traditional AI methods fall short in addressing communication security challenges due to their inability to adapt to constantly changing wireless characteristics and sophisticated cyber threats. These methods are usually trained on specific datasets, which limits their effectiveness in unfamiliar conditions and unknown attacks. Additionally, collecting enough labeled data for physical layer attacks is difficult due to the complexity and variability of noise patterns, signal interference, and channel conditions \cite{wang2023generative}. Therefore, advanced AI models are needed to learn from and adapt to these environmental factors for robust security.


Generative Artificial Intelligence (GAI) is an essential AI technology that has proven to be remarkably effective in text, images, and audio generation. 
Unlike traditional AI models, GAI operates as an unsupervised learning, which can independently identify characteristics of data and generate new samples that closely resemble the original dataset. For example, Stable Diffusion, developed by Stability AI, is an advanced text-to-image generation model, which can create highly detailed and accurate images based on textual descriptions\footnote{https://stability.ai/news/stable-diffusion-3}. 
This ability of GAI to learn, replicate, and innovate is particularly useful in addressing crucial challenges in communication security, such as handling incomplete information and imbalanced data \cite{cai2020spectrum}. Furthermore, the integration of GAI significantly strengthens the security framework, allowing for effective learning of both emerging and complex threats \cite{shi2020generative}. Despite its potential, GAI needs to be improved further due to high computational complexity and limited adaptability. For instance, Stable Diffusion was trained using 256 Nvidia A100 GPUs on Amazon Web Services for a total of 150,000 GPU-hours\footnote{https://en.wikipedia.org/wiki/Stable$\_$Diffusion}, which requires it to take relatively more computing resources and computing speed in the training and inference phases.

To address the aforementioned challenges effectively, the integration of the Mixture of Experts (MoE) approach into GAI-based models such as ChatGPT\footnote{https://medium.com/@seanbetts/peering-inside-gpt-4-understanding-its-mixture-of-experts-moe-architecture-2a42eb8bdcb3} presents a promising strategy. MoE is a machine learning ensemble technique that divides a complex task into several subtasks, which are then solved by specialized models, referred to as “experts." The MoE approach uses a gating mechanism to decide which experts to consult for a given input, effectively combining their strengths to improve overall performance on diverse or large-scale problems. 
In addition to language models, traditional AI models combined with MoE also have several applications in wireless communication security. For instance, ADMoE is an enhanced neural network-based anomaly detection algorithm by MoE \cite{zhao2023admoe}, which demonstrates superior performance in noisy label detection.

Therefore, deploying MoE-enabled GAI
models in wireless communication systems provides major advantages:

\begin{itemize}
    \item \textbf{Flexibility and Specialization:} 
    By selecting different experts for different attack types, the MoE framework ensures that the GAI model can provide more targeted and effective defense against complex attackers. For example, TB-CBR \cite{pinzon2011real} selects different experts based on the feature of attacks to assign the most appropriate techniques for identifying the type of attack.


    

    \item \textbf{Scalability and Efficiency:} 
    By strategically allocating computational resources to the most pertinent experts for a specific task or learner, MoE-based systems can simultaneously manage a larger number of experts, thereby facilitating personalized learning at scale efficiently. 
    
\end{itemize}

In conclusion, 
utilization of MoE in GAI models can enhance performance by dynamically allocating computational resources to specialized sub-models, thereby improving efficiency and adaptability in handling complex security tasks. 
Based on these, we propose a framework to utilize MoE structure in this article, enhancing the performance of GAI methods in communication security. The contributions of this paper are summarized as follows.
\begin{itemize}
    \item
    We delve into the security considerations in the physical layer and provide a detailed analysis of GAI structure in communication security. Then, we discuss the benefits of implementing GAI models and explore the applications in physical layer communication security. 
    \item 
    We present a general MoE-enabled GAI framework to solve optimization problems in communication security, which leverages the MoE's structural characteristics to enhance the protection performance of security protection algorithms in various scenarios.
    \item
    We consider friendly jamming strategies in multi-user scenarios as a case study. We employ the MoE-enabled GAI reinforcement learning algorithm to solve the optimization problem and analyze the algorithm's efficacy improvement under various indicators.

\end{itemize}

\begin{figure*}[htp]
    \centering
    \includegraphics[width= 0.9\linewidth]{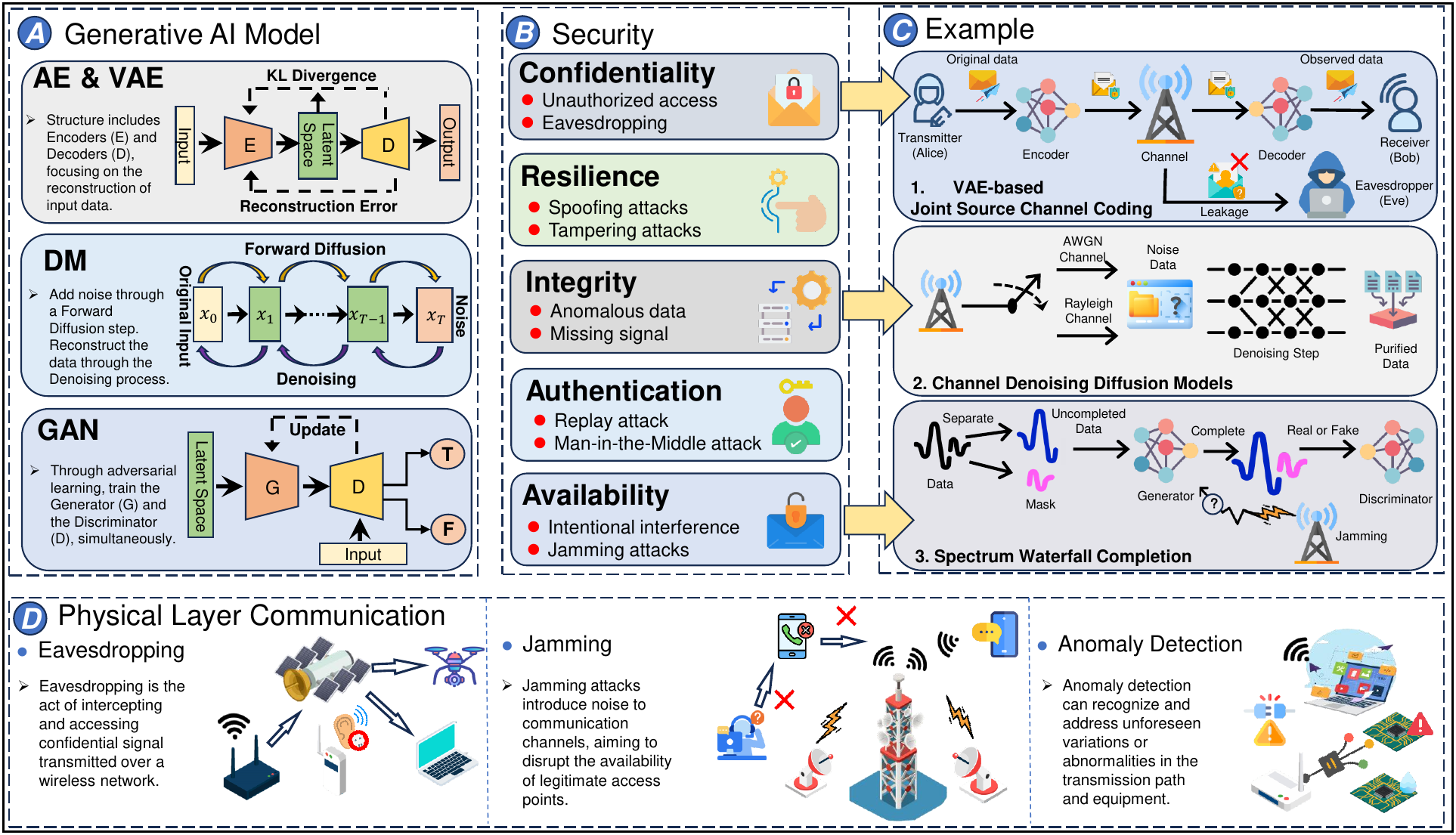}
    \caption{The overall of GAI for physical layer communication security.
    \textit{Part A} illustrates the structure of GAI models.
    \textit{Part B} presents security issues faced by physical layer communications.
    In \textit{Part C}, we introduce three examples based on GAI for communication security. These three models are based on VAE \cite{erdemir2022privacy}, DM \cite{wu2023cddm}, and GAN \cite{cai2020spectrum}, respectively.
    \textit{Part D} demonstrates important physical layer communication attack scenarios, including eavesdropping, jamming, and anomaly detection.
    }
    \label{fig:GAI}
\end{figure*}

%% file: Section2.tex
\section{Generative AI and Mixture of Experts in Physical Layer Communication Security}

This section provides an overview of GAI models, and MoE concept, exploring how GAI and MoE can be effectively employed in physical layer communication security.

\subsection{Generative AI for Physical Layer Communication Security}

GAI models, which are based on unsupervised learning, have an ability to achieve self-learning from patterns and features in the data.
They can emulate the dataset as closely as possible without needing label information. This capacity allows them to effectively address common data challenges in communication security, such as missing datasets and imbalanced data
(Figure \ref{fig:GAI}).
Currently, GAI models used in communication security include:

\begin{itemize}
    \item \textbf{Autoencoder (AEs) and Variational Autoencoders (VAEs):}
    AEs and VAEs compress the input into a lower-dimensional code and reconstruct the output from this representation as close as possible to the original input. 
    Due to its encoding and decoding structure, it can be used to ensure Confidentiality, including wiretap code design, Joint Source Channel Coding (JSCC) \cite{erdemir2022privacy}. 
    Moreover, according to the reconstruction error, they can keep signal Integrity, aiding in anomaly detection.

    \item \textbf{Generative Adversarial Networks (GANs):}
    GANs use adversarial learning to simultaneously improve the generation ability of the generator and the detection ability of the discriminator.
    Through adversarial learning, models can generate data similar to the dataset and enhance overall performance, ensuring data Availability and improving Resilience against various attacks, including spoofing attacks \cite{shi2020generative}.
    
    \item \textbf{Diffusion Models (DMs):}
    DM is a novel generative model that employs a unique technique for learning the underlying distribution of data. It first adds noise to the data and then removes it, effectively generating new data that follows the same distribution as the original. Due to its strong learning ability to added noise, DM is able to effectively learn and purify attacked data, preserving the Integrity of the data \cite{wu2023cddm}.

\end{itemize}

Even GAI models have different focuses and characteristics. It still faces some problems in dealing with today's more complex security issues.

\begin{itemize}
    \item \textbf{High Complexity:}
    GAI models are known for their sophisticated algorithms, complex training processes, and long inference times, contributing to their high complexity. When facing rapidly changing environments, they may struggle to infer and protect against cyber threats with limited resources.


    \item \textbf{Low Adaptability:}
    Due to its dependence on predefined datasets and algorithms, the inherent inflexibility can challenge maintaining effectiveness in diverse or novel contexts. 
    As the model expands, retraining it from scratch becomes increasingly challenging, rendering it less adept at responding to unexpected events or data anomalies \cite{wang2024toward}.

    \item \textbf{Limited Detection Performance:}
    GAI models often focus on one or a few specific security threats, and when faced with complex real-world tasks, they typically demonstrate limited performance. This limitation is due to their high complexity and low adaptability, which can hinder the AI's ability to accurately identify new, subtle, or sophisticated threats that differ from its training sets.
    
\end{itemize}

Given the challenges faced by existing GAI models, new architectures have been proposed to overcome these issues. MoE is an innovative approach aiming to enhance GAI's flexibility and effectiveness in dealing with a broader range of security threats by combining the strengths of multiple specialized models, i.e., experts.

\begin{figure*}[htp]
    \centering
    \includegraphics[width= 0.95\linewidth]{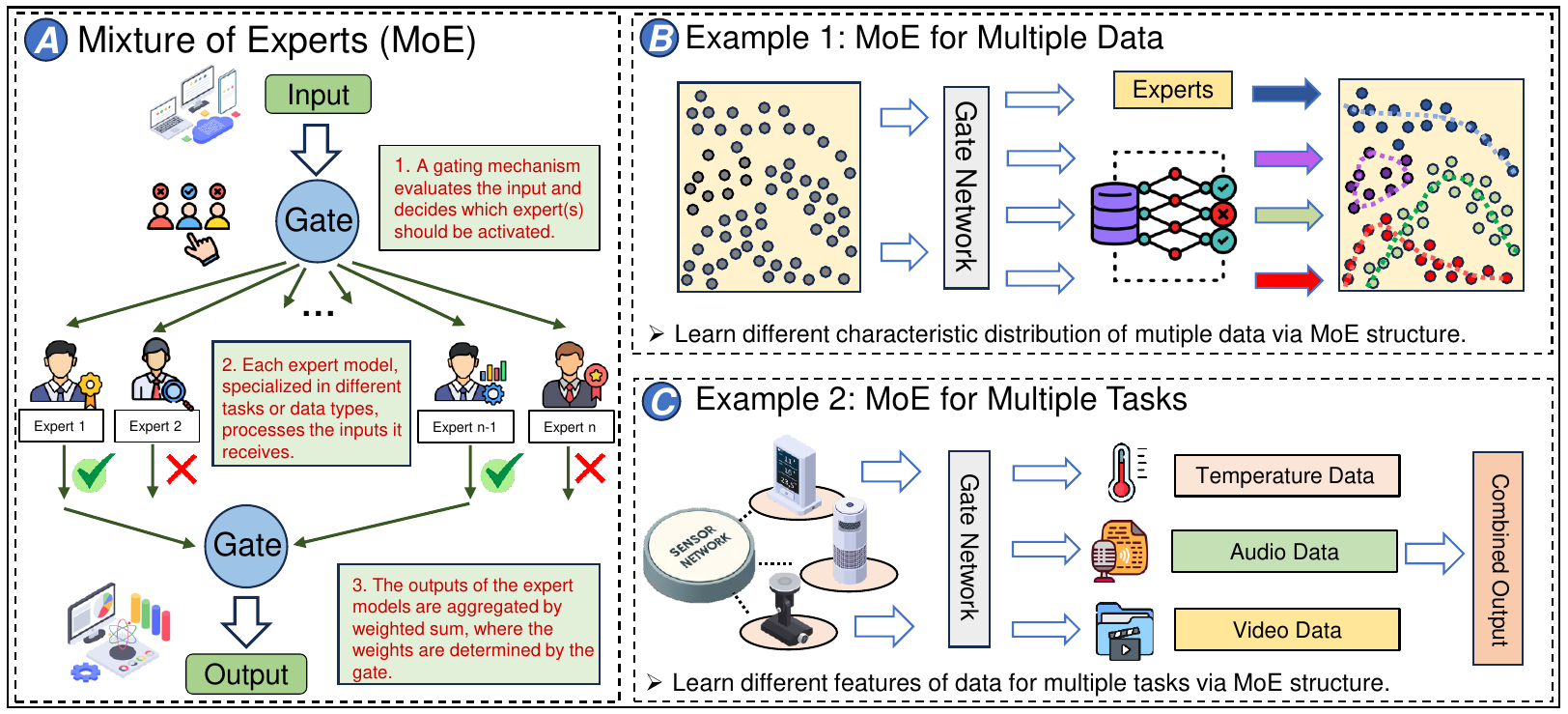}
        \caption{The structure of MoE.
    In \textit{Part A}, we illustrate the main components and working mechanism of the MoE framework.
    \textit{Part B} presents an example of using the MoE framework to learn the different characteristic distributions of multiple data.
    \textit{Part C} demonstrates how to learn different features of data for multiple tasks via MoE structure. }
    \label{fig:moe}
\end{figure*}

\subsection{Mixture of Experts}

MoE improves model performance, adaptability, and scalability by assigning tasks to specialized sub-models, or “experts," for a tailored approach to managing complex data \cite{wang2024toward}. 
By leveraging the unique strengths and expertise of multiple sub-models, each specialized in particular data types or tasks, MoE aims for higher accuracy and efficiency in computational problem-solving. 
An MoE model includes the following important components:


\begin{itemize}
    \item \textbf{Gating Mechanism:} A gating mechanism evaluates the input and decides which expert(s) should be activated for a given input, effectively routing the data to the most relevant expert models.
    \item \textbf{Expert Models:} Each expert model, specialized in different tasks, processes the inputs that it receives. This specialization allows for tailored processing and improved handling of complex patterns within the data.
    \item \textbf{Combining Outputs:} The outputs of the expert models are then aggregated. The method of aggregation can vary, including weighted sum, where the weights are often determined by the gating mechanism's output.
\end{itemize}
Figure \ref{fig:moe} presents the main components of MoE and explain its working mechanism through two examples.



Due to its dynamic allocation capabilities and efficiency in tackling complex problems, the MoE framework has become widely used in Large Language Models (LLMs). MoE enables these models to manage diverse and complex linguistic tasks more effectively. 
For instance, 
Mixtral 8x7B\footnote{https://mistral.ai/news/mixtral-of-experts}, a state-of-the-art transformer model developed on the MoE framework, has 46.7B total parameters but only uses 12.9B parameters per token. 
This efficient parameter usage enables Mixtral to process inputs and generate outputs with the speed and cost efficiency of a 12.9 billion parameter model. Notably, Mixtral achieves competitive or superior performance compared to GPT-3.5 across a range of benchmarks.

The above examples show that AI combined with MoE can bring a variety of improvements. Especially, in the field of communication security, combining MoE in GAI can bring many potential benefits:

\begin{itemize}
    \item \textbf{Enhanced Detection:}
    Learning the distribution of multiple hybrid attacks, including various spoofing attacks and jamming attacks, is usually a challenging task, which requires advanced structures for GAI models. By utilizing the MoE structure that combines multiple expert models, a single sub-model can focus on learning the distribution of certain attacks, thus improving the fitting capability for the attacks with a relatively simple model.
    
    \item \textbf{Resource Efficiency:} 
    In the inference phase, the GAI model needs to use all parameters under its complex network structure, making it difficult to deploy and perform real-time calculations on edge devices with limited resources.
    The gate mechanism enables the MoE model to activate only the most suitable sub-model for the specific task at hand, optimizing the use of available resources and ensures a more effective response to cyber threats.
    

    \item \textbf{Adaptive Learning:}
    In wireless communications, whenever the GAI model needs to learn the distribution of a new attack, such as jamming attacks from a new source, all parameters of the model need to be trained. The MoE uses a gate mechanism to train the key parameters of GAI model, reducing the training cost of the overall model, and improving the overall adaptability.

\end{itemize}


These features can address issues encountered by GAI models mentioned above, providing a foundation for using the MoE structure to improve the performance of GAI models in communication security.


\section{Applications of Generative AI and Mixture of Experts for Physical Layer Security} \label{App}





GAI technology has numerous applications for wireless communications. In this section, we present several GAI applications in physical layer communication security and provide examples of combining GAI with MoE (Table \ref{tab:data}).

\begin{table*}[htp] \scriptsize
  \centering
  \captionsetup{justification=centering}
  \caption{Summary of GAI and MoE for communication security. \\Blue circles describe the methods; Green correct markers and Red cross markers represent pros and cons respectively.}
  \label{tab:data}
    \begin{tabular}{m{0.11\textwidth}<{\centering}||m{0.06\textwidth}<{\centering}|m{0.09\textwidth}<{\centering}|m{0.64\textwidth}<{\centering}}
      \hline
      \textbf{Security Consideration}  &  \textbf{Reference} & \textbf{Algorithm} & \textbf{Pros \& Cons} \\
        \hline
        Confidentiality & \cite{erdemir2022privacy}  & VAE-based JSCC & \begin{itemize}[leftmargin=*]
        \item[\textcolor{blue-green}{\ding{108}}]
      A data-driven approach using VAE-based JSCC
          \item[\textcolor{green}{\ding{51}}] Conceal the confidential data different from the original signal
      \item[\textcolor{red}{\ding{55}}] Assume worse eavesdropping channel quality
        \vspace{-1.0em}
        \end{itemize}\\
        \hline
        \multirow{2}{0.11\textwidth}[-10pt]{\centering Integrity} & \cite{wu2023cddm}  & CDDM & \begin{itemize}[leftmargin=*]
     \item[\textcolor{blue-green}{\ding{108}}] A channel denoising diffusion models to remove noise in communications channels
          \item[\textcolor{green}{\ding{51}}] Remove the channel nosie under multiple fading channels
      \item[\textcolor{red}{\ding{55}}] Require relatively long sampling time
      \vspace{-1.0em}
      \end{itemize}\\
      \cline{2-4}
      & \cite{yu2021mixture}  & MEx-CVAEC  & \begin{itemize}[leftmargin=*]
     \item[\textcolor{blue-green}{\ding{108}}] An anomaly detection method combined with MoE framework
          \item[\textcolor{green}{\ding{51}}] Learn multiple latent space features instead of a single feature space
      \item[\textcolor{red}{\ding{55}}] Lack experiments on complex data sets
      \vspace{-1.0em}
      \end{itemize}\\
      \hline
        Availability & \cite{cai2020spectrum}  & GAN & \begin{itemize}[leftmargin=*]
        \item[\textcolor{blue-green}{\ding{108}}] A GAN-based algorithm designed to uncover the data relationships and fill in the gaps in the data
          \item[\textcolor{green}{\ding{51}}]  Recover spectrum data in multiple jamming conditions 
      \item[\textcolor{red}{\ding{55}}] Lack real-world experiments
      \vspace{-1.0em}
      \end{itemize}\\
      \hline
    \end{tabular}
\end{table*}


\subsection{GAI for Confidentiality}

JSCC optimizes both source coding (compression) and channel coding (error correction) in a unified approach to enhance efficiency and reliability of data transmission while safeguarding data confidentiality against eavesdroppers over a wiretap channel. 
GAI models, particularly AEs and VAEs, are able to encode data into a latent space and then decode it back to the original space through distribution decoding in the feature space, effectively serving as a JSCC mechanism.
For example, a VAE-based JSCC approach proposed in \cite{erdemir2022privacy} was designed to minimize information leakage and protect sensitive information from being deciphered by unauthorized parties. 
Specifically, this approach involves a transmitter (Alice) encoding source data and a receiver (Bob) aiming to reconstruct the encoded data with high fidelity, while ensuring that the sensitive information remains undetectable by any potential eavesdroppers (Eves). This is achieved by maximizing the mutual information between the user’s data and the noisy codewords observed by Bob, while minimizing the distortion to improve the pixel-wise data reconstruction quality.
According to the experiments, the proposed method can achieve good performance in both single channel and parallel channels scenarios. Especially in the parallel Channels scenario, Eve’s classification accuracy is only 20\% on average \cite{erdemir2022privacy}.

\subsection{GAI for Integrity}

During the transmission of data, it will be affected by factors such as noise and attacks, which will affect the integrity and accuracy of the data received by receivers, even with the presence of attacks to alter the data maliciously.
Due to its inherent ability to progressively remove noise, DMs have the potential to aid the receivers in mitigating channel noise and purifying received data.
A representative example is Channel Denoising Diffusion Models (CDDM) \cite{wu2023cddm} designed to leverage the noise reduction properties of DMs. The proposed model consists of three stages, where the first and last stages are the JSCC encoder and decoder trained to minimize the reconstruction error, respectively.
The second stage utilizes a noise schedule that closely simulates the distribution of channel noise,
rendering the CDDM adaptable to a variety of channel conditions.
The efficacy of integrating CDDM into communication systems is evident from the results, which show a consistent performance improvement across all signal-to-noise ratio (SNR) levels. Specifically, the inclusion of CDDM under an Additive White Gaussian Noise (AWGN) channel and a Rayleigh fading channel at 20 dB SNR results in gains of 0.49 dB and 1.06 dB \cite{wu2023cddm}, respectively.

\subsection{GAI for Availability}

The jamming attack aims to cause interference by introducing noise, thereby disrupting legitimate communications at the physical layer. Jamming attacks often result in incomplete data due to their attack characteristics, which hinders the ability of anti-jamming strategies to accurately identify jamming attacks.
The authors in \cite{cai2020spectrum} have proposed an efficient algorithm that makes use of a GAN to complete missing information in a spectrum waterfall. A spectrum waterfall is a thermodynamic block diagram that defines the environmental state. By automatically mining relationships within the data, the algorithm can accurately complete any missing data. Unlike an original GAN, where noise inputs were used, the authors utilized the spectrum waterfall with the missing data as the generator input, consequently limiting the generator's artistry. According to the corresponding complement results, the proposed algorithm outperforms the method without pre-classification since the generator adding auxiliary information to the data is more targeted. The accuracy of the proposed algorithm is over 95\%, compared to the nearly 80\% accuracy of the method that does not use pre-classification \cite{cai2020spectrum}.

\subsection{MoE-based GAI for Security}

Typically, reconstruction-based GAI models for anomaly detection only consider a single latent space. However, multiple latent spaces contain various correlation discriminative features, each contributing differently to efficient anomaly detection. 
MEx-CVAEC \cite{yu2021mixture} is an MoE-based model, comprising two convolutional VAEs and CNNs. Each component operates as an expert with an identical procedural framework to efficiently process highly relevant information. 
According to the latent space visualization comparison, a model with multiple expert structures can better cluster normal and abnormal data than a single expert. Compared with other unsupervised learning or GAI baselines, the mean Area Under the Curve (AUC) of this model based on three different datasets can achieve the best performance \cite{yu2021mixture}. In conclusion, by leveraging the potential of several latent spaces via the MoE structure, the overall detection performance can be enhanced.

\subsection{Lessons Learned}

From above applications, we can draw the following key observations:

\begin{itemize}
    \item GAI models are able to learn the distribution of the dataset and then reconstruct the data from the distribution, which is suitable for secure communications. For instance, VAE models can be utilized to improve the encoding of physical layer data transmission \cite{erdemir2022privacy}.

    \item Different from traditional AI models, GAI models can purify data by learning the characteristics of attacks or noise. In the above example, if the data is not purified, the accuracy of the received information will be seriously affected and the integrity cannot be ensured.

    \item By practicing adversarial learning, GAI has the ability to learn data features more precisely, and also enhance the accuracy of the discriminator. This approach resolves the issue of data incompleteness and imbalance caused by intricate scenes such as jamming attacks.
    

    \item The MoE structure allows AI detection models to utilize more complex data including multi-label, multi-category, and multi-latent distributions, improving detection accuracy compared to a single model.
\end{itemize}


However, most existing works only
consider certain scenarios and lack real-world testing, since they have difficulty to infer and adapt to additional fast time-varying
information in real-time \cite{cai2020spectrum}.
Therefore, using the MoE structure has become one of the feasible solutions to solve GAI's issues in communication security.


%% file: Section3.tex
\section{MoE-enabled Generative AI for Physical Layer Communication security}

In this section, we discuss research challenges in communication security optimizations.
Then, we propose an MoE-enabled GAI framework focusing on optimization problems in communication security.


\subsection{Research Challenges}

Communication security emphasizes the design of secure, safe, and reliable information transmission methods \cite{wang2018survey}. From the perspective of system optimization and design, ensuring communication security mainly includes the following aspects:



\begin{itemize}
    \item \textbf{Resource Allocation:}
    Secure resource allocation aims to optimize limited communication resources, such as bandwidth and energy, to maximize the requirement of various performance metrics, including 
    Secrecy Rate (SR), Secrecy Outage Probability (SOP), power consumption, and Energy Efficiency (EE) \cite{wang2018survey}.
    
    

    
    \item \textbf{Signal Processing:}
    Secure beamforming and precoding are advanced signal processing techniques designed to enhance the confidentiality of wireless communications. By intelligently shaping and directing transmitted signal beams, these methods minimize potential eavesdropping and jamming, thus optimizing the network's SR and EE.

    \item \textbf{Antenna Selection and Cooperation:}
    In multi-antenna wireless networks, selection and cooperation of antennas can improve performance metrics such as secrecy rate and outage probability \cite{du2023beyond}. These strategies can reduce power consumption while boosting EE.
\end{itemize}

Reinforcement Learning (RL) algorithms based on DMs have recently achieved success in solving such communication optimization problems, including
UAV path planning and resource allocation \cite{du2023beyond}.
However, it usually focuses only on a single optimization goal and cannot address joint considerations of communication security \cite{wang2018survey}:

\begin{itemize}
    \item [-] The transmission effectiveness is evaluated through the achievable SR.
    \item [-] Reliability is measured in terms of SOP.
    \item [-] Power cost refers to the minimum power consumption required to ensure secure Quality of Service (QoS).
    \item [-] EE is concerned the quantity of secret bits transferred per unit of energy.

\end{itemize}

\begin{figure*}[htp]
    \centering
    \includegraphics[width= 0.95\linewidth]{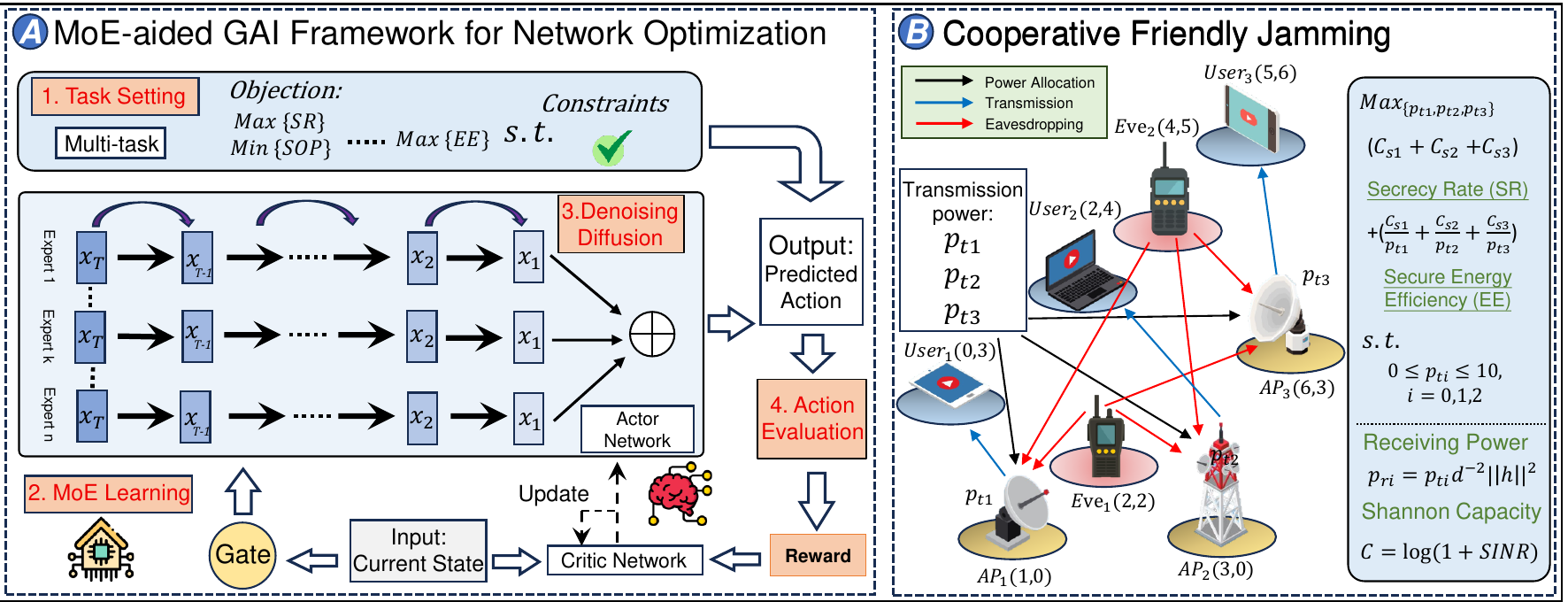}
    \caption{The workflow of the proposed framework. 
    In \textit{Part A}, the proposed framework solves optimization problems through reinforcement learning following 4 steps: 1). Task Setting; 2). MoE Learning; 3). Denoising Diffusion; 4). Action Evaluation.
    \textit{Part B} illustrates the cooperative friendly jamming optimization problem considered in Section \ref{Case}. The figure introduces the detail scenario and optimization problem settings including optimization objectives, constraints, etc.}
    \label{fig:frame}
\end{figure*}

\subsection{Framework Design}

Based on the characteristics of MoE and GAI framework, in this part, we propose the MoE-enabled GAI framework to solve communication security optimization problems.

\subsubsection{Architecture}

Our framework follows the GAI-based RL model, GDM \cite{du2023beyond}. To consider multiple performance metrics at the same time, we incorporate the MoE structure after the state input layer. By combining the MoE structure, different experts can focus on a few performance metrics with a shared-bottom model \cite{ma2018modeling}, rather than optimizing all metrics at the same time. For instance, suppose two performance metrics need to be optimized in a wireless communication system, including secrecy rate and energy efficiency. In such a case, the GAI framework with MoE structure can be trained by combining two experts. Each expert can be automatically assigned to capture either shared performance metrics information or a specific one, which 
is more conducive than one network to finding the theoretically optimal solution for two metrics.  



Specifically, regarding secure communication, the definitions of the state space, action space, and reward are as follows:



\begin{itemize}
    \item \textbf{State Space:} 
    The state space consists of the current information and previously selected actions.
    The state is the channel information, which includes characteristics such as channel quality, current interference levels, signal fading, and other relevant environmental conditions affecting signal transmission. 

    \item \textbf{Action Space:}
    The action space encompasses a range of adjustable parameters that can significantly influence its performance and security. Examples of these parameters include frequency, bandwidth, transmission power, and devices distance. 


    \item \textbf{Reward:}
    The reward based on each action taken at the current state is determined by security performance metrics \cite{wang2018survey}. 
    The performance metrics are different security metrics including 
    SR, SOP, and EE, etc.

    

\end{itemize}

\subsubsection{User Workflow}

In the proposed framework, to solve an optimization problem about communication security, one may follow these steps (Figure \ref{fig:frame} \textit{Part A}):


\begin{itemize}
    \item \textbf{Task Setting:} 
    To optimize a secure transmission, users must first confirm reasonable performance metrics that consider security performance. This is usually achieved through a variety of metrics, which are then divided into individual tasks. For example, when designing a system to prevent eavesdropping, users can consider the hybrid metrics.

    \item \textbf{MoE Learning:} 
    By using the gate mechanism in MoE, the input states can be divided into individual independent expert models. Then different experts can automatically capture 
    either shared and task-oriented information, allowing them to obtain higher action reward.
    For instance, users can assign three or more experts to optimize different performance metrics, through the learning of gate network.
    


    
    \item \textbf{Denosing Diffusion:}
    Based on the GDM structure \cite{du2023beyond}, users predict actions in the current state by using conditional denoising to filter out noise. The conditions used in this process can correspond to various subtasks, such as current state, previously selected action, or a combination of both. For instance, when focusing on improving security performance, the users can choose the current channel state and the previous transmission power as conditions in the subtask of transmission power.

    
    \item \textbf{Action Evaluation:}
    Instead of denoising towards the maximum value of the probability distribution, the actor network is learned by denoising towards the direction that maximizes reward, and then combined with the gate network to obtain the finial decision.
    
\end{itemize}

Combining the powerful generation capability of GAI with learning for an optimal action denoising can help to explore the optimal solution more effectively \cite{du2023beyond}. By incorporating the MoE structure, the framework can focus on multiple performance metrics simultaneously, achieving better results in multi-objective optimization problems.

%% file: Section4.tex
\section{Case study: Cooperative Friendly Jamming for Physical layer Communication security}
\label{Case}

In this section, we present a case study that demonstrates the optimization for cooperative friendly jamming and utilize the proposed framework to demonstrate how MoE can improve the performance of GAI models.

\subsection{System Model}

We consider a cooperative friendly jamming scenario, where Access Points (APs) produce friendly jamming signals to further degrade eavesdropping \cite{hoseini2024cooperative}. 
This scenario involves multiple APs, multiple users, and potential eavesdroppers. When a user downloads data from a certain AP, the other APs will act as jammers to prevent any eavesdroppers from intercepting the user's traffic.
In this case study, we consider 3 APs, 3 users, and 2 eavesdroppers, where each user communicates with one AP (Figure \ref{fig:frame} \textit{Part B}). 
For the received power $p_r$ of the system, we consider large-scale fading based on distance and small-scale fading based on the Rayleigh fading model.
For a given user or eavesdropper, the Shannon capacity is determined by the Signal to Interference plus Noise Ratio (SINR), where SINR is calculated from the received APs power and noise.




\subsection{Optimization Problem Formulation}



We aim to jointly maximize SR and secure EE \cite{wang2018survey} to find the optimal power allocation to the APs.
The effectiveness of physical layer communication security strategies can be measured using SR. Our assessment of SR involves comparing the data rate of the legitimate channel to that of the wiretap channel with the Gaussian codebook.
For a given user, the SR $C_s$ is calculated via the Shannon capacity minus the largest Shannon capacity of eavesdroppers.
Our goal at EE is to ensure that the transmission strategy operates in a confidential and eco-friendly manner. To achieve this, we utilize secure EE, which is the amount of secret bits transmitted with the consumption of one unit of energy.
For a given AP, its secure EE $S_e$ is determined by the SR of the corresponding user to AP's transmission power $p_t$.

\subsection{Numerical Results}

\begin{figure}[!]
\centering
\begin{subfigure}{\linewidth}  
\centering
\includegraphics[width=0.85\columnwidth]{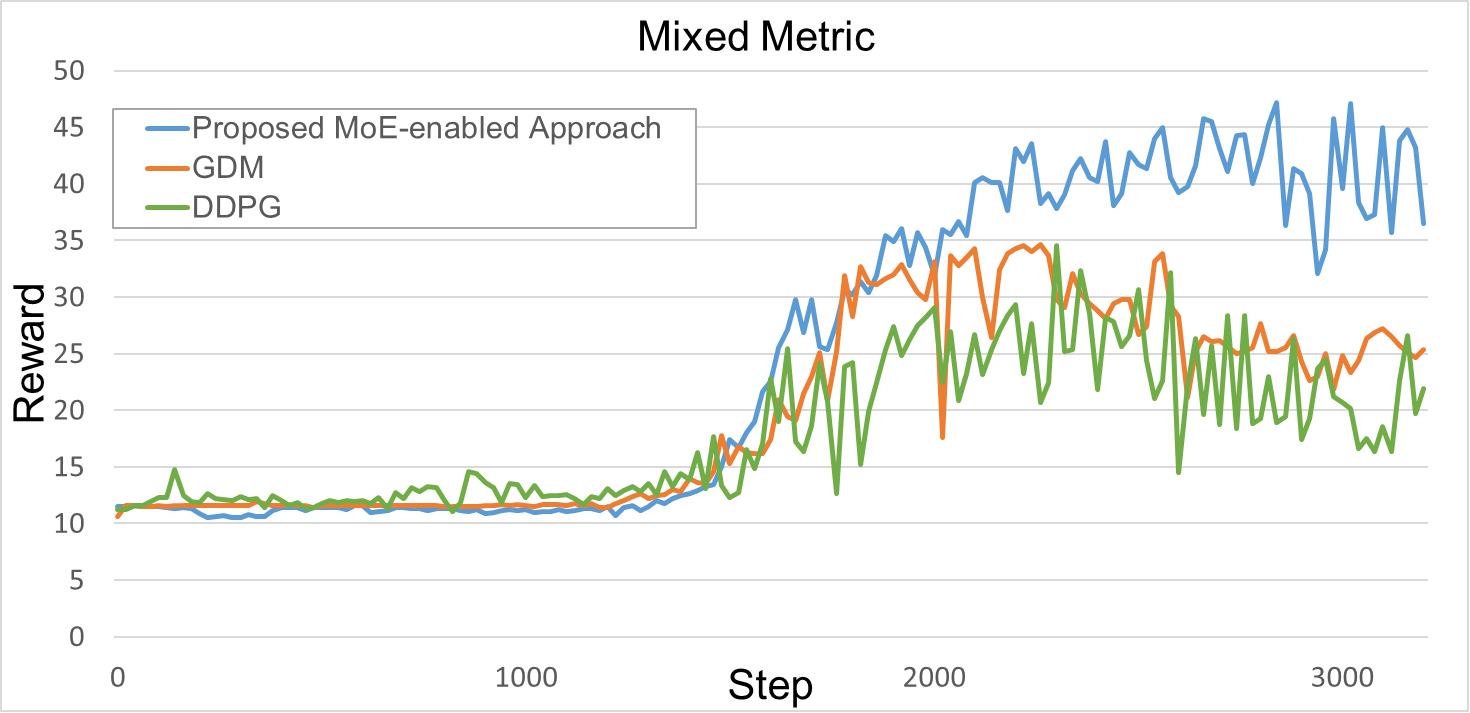}
\caption{Sum of SR and secure EE comparisons for three methods.}
\label{fig:mix}
\end{subfigure}
\begin{subfigure}{\linewidth}  
\centering
\includegraphics[width=0.95\columnwidth]{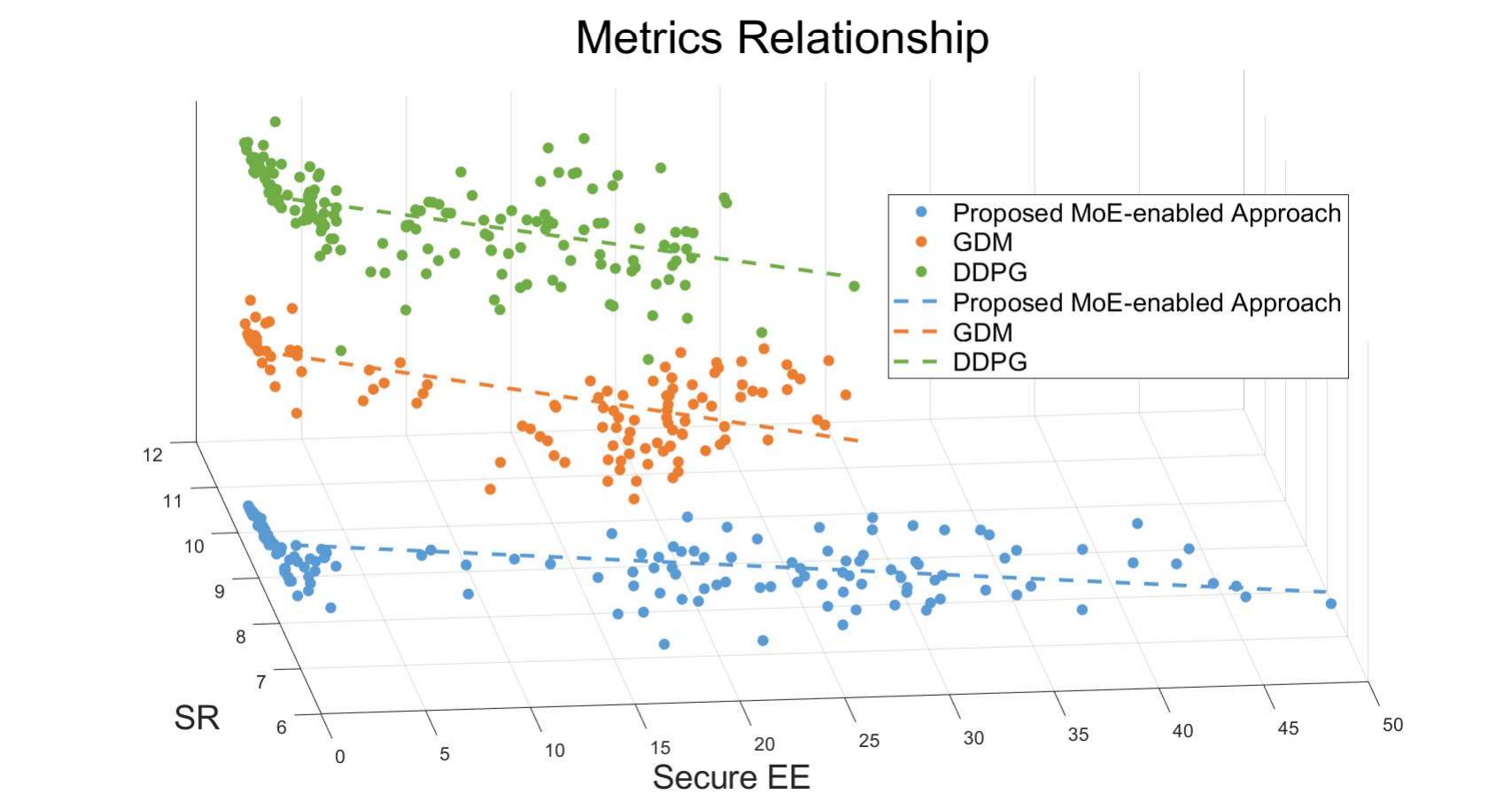}
\caption{Comparison of metrics relationship for three methods.}
\label{fig:sr}
\end{subfigure}
\caption{Performance comparison of three methods in cooperative friendly jamming scenario.}
\label{fig:main}
\end{figure}

\subsubsection{Experimental Setup}

Our experiment is based on Python along with the PyTorch package, conducted on a V100 Linux server. The main program is based on the GDM framework \cite{du2023beyond}. The proposed MoE method includes 3 experts and selects the best expert each time.

\subsubsection{Performance Analysis}

Figure \ref{fig:main} illustrates the comparison of various performance metrics between the proposed MoE-enabled GAI method, GDM without MoE, and Deep Deterministic Policy Gradient (DDPG) algorithm \cite{du2023beyond} in a given scenario. Among them, the blue curve represents the MoE method, the orange curve represents GDM, the green curve represents DDPG,
and the reward represents the specific numerical value of metrics. 
Based on the transmission power derived from two approaches, Figure \ref{fig:mix} illustrates the learning curve for the sum of SR and secure EE. Clearly, the proposed method demonstrates the ability to converge to a reward of approximately 40.56
and has the potential to explore and achieve a reward as high as 47.19. 
In comparison, the reward obtained through GDM tends to converge around 27.99, with the maximum reward during training reaching only 34.63. The DDPG method performs even worse, achieving only around 23.17 and the maximum reward capped at 34.52.
Figure \ref{fig:sr} shows the numerical relationship between the two indicators among the three methods. For a more linear fitting trend line, as Secure EE decreases, both SR metrics show a downward trend. However, the proposed method (organe curve) decreases slowly, and higher Secure EE values can be explored compared to the other two methods with the same SR value.






\subsection{Discussion}


The experimental results above demonstrate that in a cooperative friendly jamming scenario, the MoE-enabled GAI optimization algorithm can achieve better exploration performance and convergence results in 
optimizing mixed metrics, including SR and secure EE.
Through the comparison results (Figure \ref{fig:main}), the MoE can explore more suitable resource allocation strategies to improve the performance of one task (secure EE) without compromising the performance of another task (SR) significantly. 
Since the MoE only activates one expert through the gate network for training and inference each time, the overall computational complexity is similar to that of the original GDM and does not increase significantly.


    

%% file: Section5.tex
\section{Future Directions}


\subsection{Zero-trust Physical Layer Communication}



Zero-trust communication refers to a security strategy where every communication attempt
must be verified.
The MoE-based GAI algorithm allows for the simultaneous maximization of various security performance metrics during the network design process, such as MoE-based DMs for graph generation.
Therefore, all network entities can be verified and trusted, minimizing the risk of tampering or interception.

\subsection{Real-time Anomaly Detection}

The high communication speed has put forward requirements for real-time anomaly detection.
By leveraging MoE structure, only important network levels will be activated according to the gate mechanism, reducing the inference speed of GAI models. 
This approach could also significantly improve the detection of sophisticated cyber-attacks that conventional systems might overlook.


\subsection{Enhanced Privacy Preservation}

Data privacy protection is also an important part of physical layer security. During transmission, MoE-enabled GAI model can process signals in a way that maximizes data privacy.
For instance, experts within the MoE model could specialize in different privacy-preserving techniques such as data anonymization, encryption, and differential privacy. 


%% file: Conclusion.tex
\section{Conclusions}\label{Conclu}




The paper demonstrated how GAI models can improve communication security and explores their capabilities when combined with the MoE framework.
Specifically, 
we started by introducing the GAI technology and its application in physical layer communication security. We emphasized superior capabilities of GAI and compared them to conventional AI models in communication security by providing examples.
Subsequently, we highlighted the limitations of the GAI model. By introducing the structure and characteristics of MoE, we explored the possibility of the MoE framework to overcome limitations of GAI models.
Finally, we proposed an MoE-enabled GAI framework focusing on network optimization problems in communication security. 
Through a case study in a cooperative friendly jamming scenario, it illustrated how the MoE structure improves GAI algorithms to ensure safety in wireless communication systems.